\documentclass[a4paper,13pt]{extarticle}
\usepackage{setspace}
\usepackage{color}
\usepackage{graphicx,graphics}
\usepackage{hyperref}
\usepackage{pdflscape}
	\usepackage{amsmath}
\title{$AstroSat$  $-$ a multi-wavelength astronomy satellite}
\author{A. R. Rao, K. P. Singh\\
Tata Institute of Fundamental Research, Mumbai, India\\
D. Bhattacharya\\
 Inter University Center for Astronomy \& Astrophysics, Pune, India}

\begin{document}
\maketitle

\begin{abstract}
 
  $AstroSat$ is a multi-wavelength  astronomy satellite,  launched on 2015 September 28.  It  carries a suite of scientific instruments for multi-wavelength observations of astronomical sources.  It is a major Indian effort in space astronomy and the context of $AstroSat$ is
  examined in a historical perspective. The Performance Verification phase of $AstroSat$ has been  completed and 
  all instruments are   working flawlessly and as planned. Some brief highlights of the scientific results are also
  given here.

\end{abstract}
{\bf keywords}: Astronomy: general,  Astronomy: instrumentation 

\section{Introduction}
   $AstroSat$, India's first dedicated astronomy satellite, was launched on 2015 September 28.  It was the 30$^{th}$  successful launch of India's workhorse rocket, the Polar Satellite Launch Vehicle (PSLV). The satellite was placed precisely in a near-Earth orbit of 650 km at 6$^\circ$ inclination, thus saving the onboard fuel meant for orbit correction for any future eventualities and ensuring a very long orbital life for the satellite. $AstroSat$, weighing 1550 kg, carries a suite of scientific instruments for multi-wavelength observations of astronomical sources (Singh et al. 2014).  Within six months of operation, the Performance Verification phase has been completed and a very complex satellite like AstroSat  is working flawlessly and as planned.

\begin{figure}
\centerline{\includegraphics[width=12.5cm]{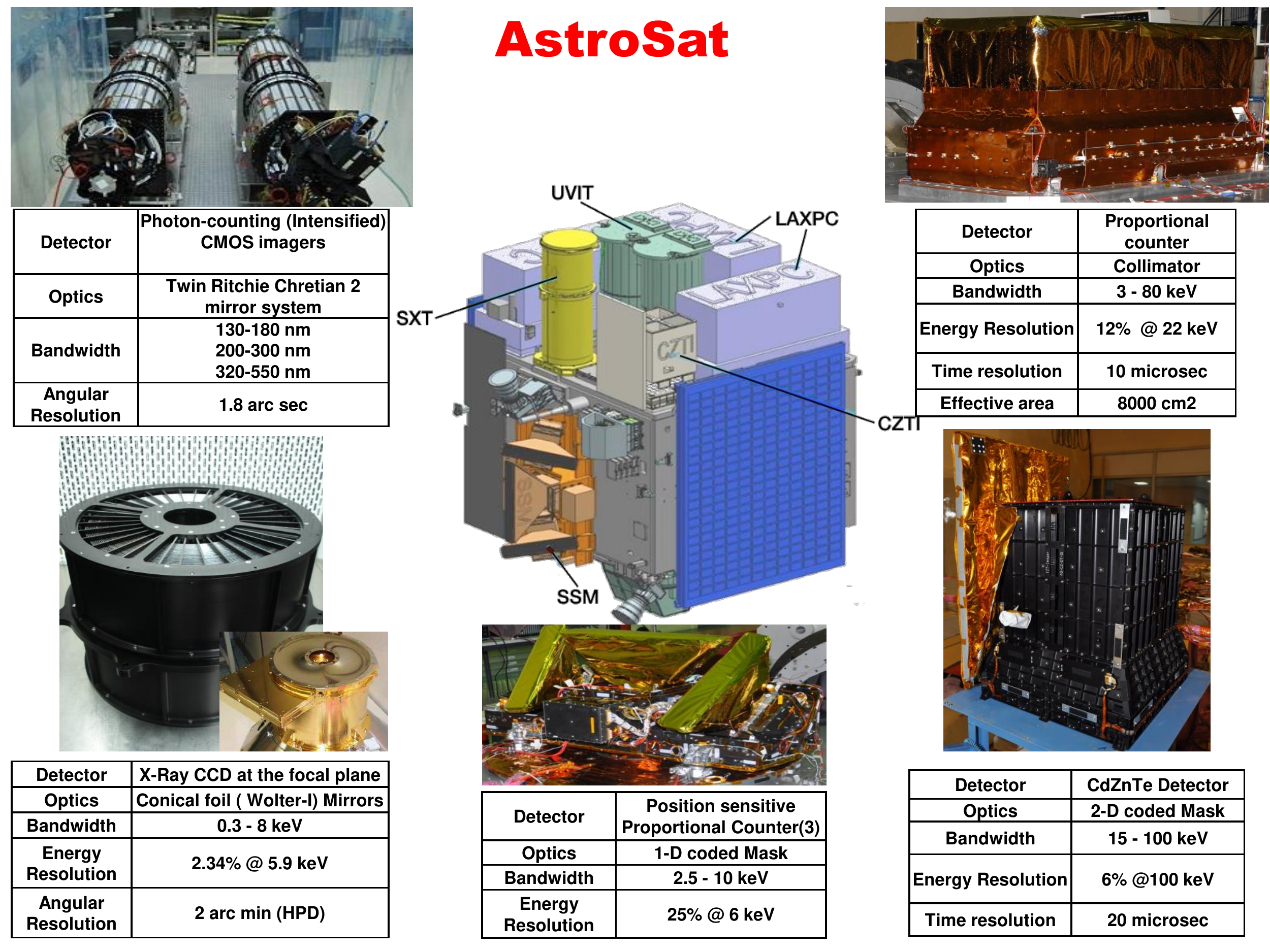}  }
\caption{An assembled view of AstroSat. The pictures of the scientific instruments  shown are (clockwise, from top left): UVIT telescope; one assembled LAXPC unit; final assembled CZT Imager; assembled SSM units; SXT mirror and the gold coated foils shown in the inset. The salient parameters of each instrument are shown in the accompanying boxes.}
\end{figure}

\section{That's one giant leap for India . . . . }

India started as a significant player in space astronomy, particularly in the exciting discovery era of X-ray astronomy in the sixties and seventies (Agrawal 1998). The country quickly converted its know-how of conducting Cosmic Ray experiments at balloon altitudes into a vibrant balloon borne hard X-ray platform (Sreekantan 2000) and also conducted some rocket-borne experiments in soft X-rays.  In the eighties and nineties, the era in which rapid strides were made in space astronomy culminating in the Great Observatories and other sophisticated satellites from NASA, ESA and Japan, India's efforts were modest in nature: Stretched Rohini Satellite Series (SROSS) detected ~60 gamma-ray bursts and the Indian X-ray Astronomy Experiment (IXAE) provided X-ray light curves of many bright Galactic X-ray binaries.

$AstroSat$ was conceived, at the turn of the century, when $Hubble$ and $Chandra$ were providing pretty pictures of the distant cosmos on a daily basis, when GRBs were firmly established to be of cosmological origin, when Rossi X-ray Timing Explorer ($RXTE$) was using the newly established web services for enlisting the expertise of the whole world to measure the `pulse of the universe'.  $AstroSat$ aims to provide reasonably good sensitivity across a wide bandwidth in the X-ray region, coupled with a precise measurement in the ultra-violet region (Figure 1). The Large Area X-ray Proportional Counters (LAXPCs) are designed to make high time resolution observations and can extend the rich legacy of $RXTE$-PCA for bright X-ray binaries and enhance the effectiveness by extending the sensitivity to higher energies; the Cadmium-Zinc-Telluride Imager (CZTI) with a Coded Aperture Mask (CAM) will extend the bandwidth to even higher energies and has the capability to provide X-ray polarization measurements in the 100  $-$ 300 keV region; the Soft X-ray Telescope (SXT) using conical-foil mirrors and an X-ray CCD provides complementary observations at low energies with very high sensitivity. The Scanning Sky Monitor (SSM) with three sets of proportional counters each equipped with 1-D CAM can continuously keep track of transients, and the Ultra Violet Imaging Telescope (UVIT) will help measure the spectral energy distributions in an extremely wide bandwidth of electromagnetic radiation. Other special features of $AstroSat$ that make it a good force multiplier are: low background due to low inclination (6 degrees), continuous time-tagging of individual photons to a few tens of micro-second accuracy (LAXPC, CZTI, and SSM), bright source observing capability and large field of view of SXT, and the facility to change/ adjust observation time of SSM.

\begin{figure}
\centerline{\includegraphics[width=12.5cm]{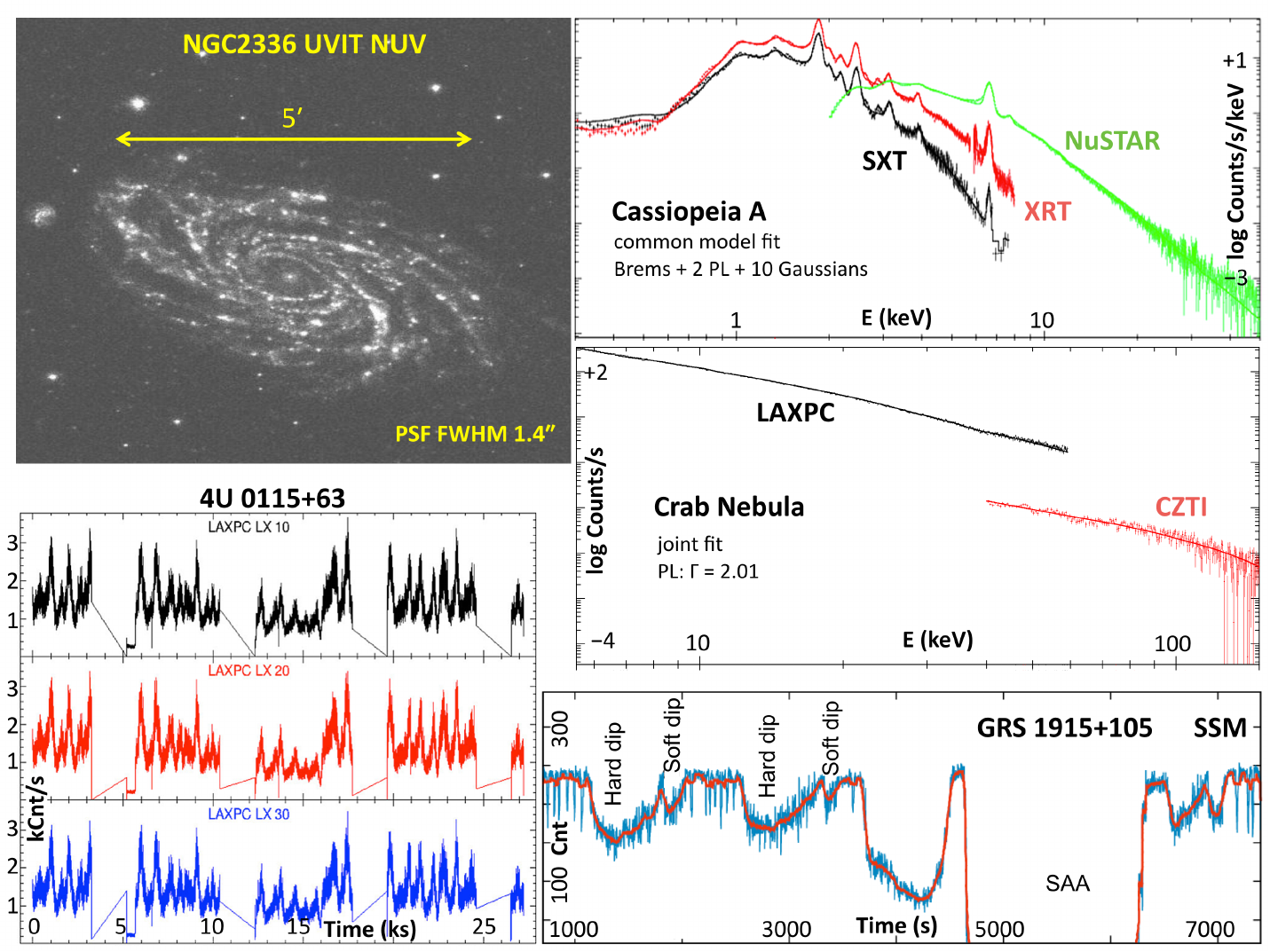}  }
\caption{Top left: UVIT image of NGC 2336 in the NUV band, demonstrating very good angular resolution. Bottom left: Light curve of 4U0115+63 during outburst in the three units of LAXPC. Top Right: X-ray spectrum of Cas A with SXT which has a spectral resolution of 140 eV; the source was observed at a large off-axis angle.   Right Middle: Joint fitting of the spectrum of Crab with LAXPC and CZTI showing simultaneous broadband spectral coverage.  Bottom Right: Light curve of GRS 1915+105 from SSM showcasing its ability to stare at a source.
}
\end{figure}

   The major technological challenge of $AstroSat$ was to make the complex instruments work reliably and without failure. The dedicated work of Indian space organization in encouraging space industries in India and developing several technologies in-house gave a tremendous boost to this effort.  Though the ability to make sealed proportional counters was demonstrated for the IXAE experiment, making a set of extremely large area deeper detectors filled at higher pressures (so that the highest ever effective area in the 10 $-$ 80 keV region is achieved) was a technological challenge. The principles of X-ray focusing technique are well known: but to make mirrors with replicated gold surface and painstakingly assemble hundreds of them at their assigned precise locations and ensuring that, all these focus the X-rays onto a small spot was extremely challenging. To ensure that sixteen thousand individual pixels of CZT detectors are well calibrated and behave in tandem to produce a hard X-ray image of the sky was a culmination of diverse sophistications in electronics design, mechanical fabrication, and onboard software. The humongous effort required for the calibration and testing of UVIT and the far-sighted dogged efforts required to achieve low contamination and sub-arc second pointing precision was truly an outstanding effort.

\section{ . . . And a significant step for mankind}

    X-ray astronomy has seen a tremendous but somewhat skewed development over the period of the past five decades. The soft X-ray regime (0.1  $-$ 8 keV) has seen a mind boggling development of a billion times improvement in the detection sensitivity, and in the medium energy range (8  $-$ 20 keV) large area detectors like $Ginga$ and $RXTE$ have explored the timing dimensions, but the detection and spectral sensitivity in this energy band got a boost only recently by NuSTAR that has pushed the focusing ability to higher energies. Still harder X-rays are a difficult regime: in the extreme hard X-rays ($>$ 80 keV), the improvement as compared to $HEAO~A-4$ (which detected a sum total of 22 sources above 80 keV) is at best marginal ($Swift$-BAT and Integral detecting about 86 and 132 sources, respectively). The impact of this skewed development in High Energy Astrophysics is profound: though we know the environments of sources of high energy emission reasonably well, the physics of high energy emission, for example, what happens close to a black hole, nature of emission from magnetars, and prompt emission of gamma-ray bursts are very poorly understood. From a handful of sources with hard X-ray timing and low energy spectral measurements in the $RXTE$ era, we now move into an era where such information will be routinely available with SXT and LAXPC. In the context of accretion onto supermassive black holes, it has been realized that to have a good understanding of the accretion phenomena, one needs measurements of reverberation between X-rays and UV/ optical region: the co-aligned X-ray and UV telescopes in AstroSat are ideally suited for making these observations.

   In this context, $AstroSat$ is a significant and important step forward, and the initial observations with $AstroSat$ have indeed shown glimpses of fructification of these promises (Figure 2). Successful tracking and compensation of arcmin-scale satellite drift and jitter has enabled the UVIT to achieve an imaging resolution of up to 1.7 arcsec.  Observations of star clusters with the UVIT have captured variable stars, including RR Lyrae types (Annapurni et al. 2016).  In the X-ray band, periodic and quasi-periodic variability has been recorded for a number of compact star binaries, using the LAXPC (Yadav et al. 2016), the CZTI as well as the SSM instruments. The SXT has demonstrated its spectral capability by clear observation of narrow spectral lines in supernova remnants such as Cas A and Tycho (Singh et al. 2016).  Broad-band continuum spectroscopy has been performed with multiple instruments, such as the LAXPC and the CZTI for the Crab Nebula.  Timing of the Crab pulsar has been carried out in detail.  The CZT imager has proved to be a capable all-sky detector of Gamma Ray Bursts. In the case of one bright GRB, hard X-ray polarization has been detected by the CZTI.  Polarization of the Crab Nebula, too, has now been measured by the CZTI at energies above 100 keV (Vadawale et al. 2016).

     Currently, concentrated efforts are being made to enhance the effectiveness of $AstroSat$.  It is envisaged that the observation program would be nimble, flexible, and responsive so that the good sensitivity of $AstroSat$ in an extremely wide range of electromagnetic radiation can be effectively used for new astrophysical phenomena like transients and varying sources in conjunction with other ground and space based observatories. Another important aspect is a detailed and precise onboard calibration, particularly for the X-ray instruments. Though the individual X-ray instruments have been calibrated as per the observing program, a detailed understanding of the systematics involved will be clearer by a more careful analysis of the simultaneous data from its various instruments. Currently, efforts are on to tune the calibration of the X-ray instruments and get well calibrated astrophysical data. It is hoped that $AstroSat$ will provide good quality wide band astronomical data for several years, hopefully partially alleviating the recent tragic loss of the $Hitomi$ satellite.


\section*{Acknowledgements}

$AstroSat$ is a project fully funded and managed by Indian Space Research Organization and realized by a consortium of a large number of Indian scientific Institutes. The authors thank the $AstroSat$ team members for providing some of the material presented in this article.

\section*{References}
 
\parindent-8pt

$~$

Agrawal, P.C., 1998, Proceedings of Indian National Academy, Vol 64, No. 3, pp407-421

	Annapurni, S. et al. 2016, \emph{In-orbit performance of UVIT on AstroSat},  Proc. SPIE Space Telescopes and Instrumentation 2016: Ultraviolet to Gamma Ray, 9905-47

	Sreekantan, B.V., 2000, in \emph{Astronomy in India: a historical perspective} Ed. T. Padmanabhan, INSA, pp61-72

	Singh K.P. et al. 2014,  \emph{AstroSat mission}, Proc. SPIE 9144, Space Telescopes and Instrumentation 2014: Ultraviolet to Gamma Ray

	Singh, K.P. et al. 2016, ``In-orbit performance of SXT aboard AstroSat’’ Proc. SPIE Space Telescopes and Instrumentation 2016: Ultraviolet to Gamma Ray, 9905-46

	Vadawale et al. 2016,  \emph{In-orbit performance AstroSat} CZTI Proc. SPIE Space Telescopes and Instrumentation 2016: Ultraviolet to Gamma Ray, 9905-48

	Yadav, J. S. et al. 2016, \emph{Large area x-ray proportional counter (LAXPC) instrument onboard AstroSat} Proc. SPIE Space Telescopes and Instrumentation 2016: Ultraviolet to Gamma Ray, 9905-45

\end{document}